# Filosofía y relatividad*

## Olimpia Lombardi.
CONICET - Universidad de Buenos Aires

*Con su Teoría de la Relatividad, Einstein produjo una profunda revolución en nuestro modo de concebir la realidad y el conocimiento que de ella podemos obtener. Esta revolución puede abordarse desde la filosofía, como conduciendo a una de las grandes cosmovisiones de la historia del pensamiento que, junto con la aristotélica y la newtoniana, marcaron las sucesivas concepciones acerca del universo y de nuestro modo de acceder a él. La comparación entre estas tres cosmovisiones nos permite comprender el progresivo descentramiento que ha sufrido el sujeto en cuanto a su posición en el cosmos y a su conocimiento de lo real.*

Suele afirmarse que, con su Teoría de la Relatividad, Einstein produjo una profunda revolución en nuestro modo de concebir la realidad y el conocimiento que de ella podemos obtener. Efectivamente así fue, y tal vez una de las mejores maneras de comprender esta revolución es abordándola desde el punto de vista de la filosofía, una disciplina que, a través de su larga historia, se ha preguntado sistemáticamente tanto acerca de la posición del ser humano en la realidad como de la naturaleza y las limitaciones de nuestro conocimiento.

Mi propósito aquí es presentar el mundo que nos describe la Teoría de la Relatividad como una de las grandes cosmovisiones de la historia del pensamiento que, junto con la aristotélica y la newtoniana, marcaron las sucesivas concepciones acerca del universo y de nuestro modo de acceder a él. Es precisamente a partir de la comparación entre estas tres cosmovisiones que podremos reconocer la verdadera dimensión de la revolución einsteniana.

Para Aristóteles, la palabra '*logos*' poseía un significado múltiple. En su acepción primaria, 'logos' significaba 'palabra' o 'discurso'; en este sentido, el logos era el principio inteligible del decir, la racionalidad misma del lenguaje. A su vez, 'logos' significaba 'concepto' o 'razón' en tanto estructura del pensamiento humano. Pero 'logos' también debía entenderse como 'ley' o 'principio', esto es, como el orden inmanente de lo real. Sin embargo, no se trataba de significaciones diferentes puesto que, para Aristóteles, la realidad, el pensamiento y el lenguaje compartían una misma estructura racional: el logos era, entonces, la racionalidad intrínseca a lenguaje, pensamiento y realidad; ello explicaba que pudiéramos alcanzar el conocimiento de lo real y expresar lingüísticamente tal conocimiento.

¿Cómo concibe Aristóteles esa realidad acerca de la cual podemos pensar y hablar? La realidad aristotélica es un cosmos, un universo ordenado, regido por leyes definidas. Pero en este universo pueden distinguirse dos regiones de diferente naturaleza: el mundo supralunar, que es el mundo de los astros, compuesto de éter, perfecto e inmutable, y el mundo sublunar, compuesto de los cuatro elementos −aire, agua, tierra y fuego−, corruptible y en permanente mutación. Cada una de estas dos regiones posee sus propias leyes. En el mundo supralunar rige la ley del movimiento circular uniforme, eterno y perfecto como el círculo mismo, el movimiento que menos parece movimiento. En el mundo sublunar, por el contrario, los cuerpos se mueven buscando su lugar natural, lugar que depende de las proporciones de los cuatro elementos que intervienen en su composición: la tierra y el agua se mueven hacia el centro de la Tierra, mientras que el aire y el fuego se mueven alejándose del centro hacia los confines del mundo sublunar.

El universo aristotélico es finito, cerrado y esférico, limitado por la esfera de las estrellas fijas, telón de fondo salpicado de puntos luminosos que contiene todo lo real. Más allá de la esfera de las estrellas fijas no hay nada: ni materia, ni espacio. Estrictamente, no tiene siquiera sentido preguntarse por un 'más allá' de la esfera de las estrellas fijas pues la realidad toda está contenida en ella.

En este universo esférico, la Tierra ocupa el centro, no sólo desde un punto de vista geométrico sino principalmente dinámico, pues el centro de la Tierra −y, por tanto, del universo− define tanto las direcciones de los movimientos sublunares, de los objetos hacia su lugar natural, como los movimientos circulares supralunares al establecer su punto central. Sin embargo, podríamos preguntarnos: la Tierra ¿es el centro del universo o está ubicada en el centro del universo? En otras palabras, el punto que define los dos tipos de movimiento, ¿es el centro físico del objeto Tierra o es el lugar, el punto central del universo esférico? Aristóteles se formula explícitamente esta pregunta y brinda una respuesta clara: el centro dinámico del universo es su centro geométrico, independientemente de estar ocupado o no por la Tierra. Incluso imagina la posibilidad de que la Tierra se desplazara de su posición actual: en este caso, volvería al centro geométrico del universo, ubicándose en su lugar natural dado por su composición, en la que prevalece el elemento tierra. Aristóteles también explica la forma esférica de la Tierra mediante la ley de los movimientos sublunares: la esfera es la forma en que el elemento tierra puede ubicarse, como un todo, más cerca de su lugar natural. Esta respuesta muestra la estrecha relación entre física y astronomía en el sistema filosófico aristotélico: los hechos astronómicos se explican por causas físicas dependientes de las leyes de movimiento.

En definitiva, el universo aristotélico es un universo esférico, no-homogéneo y no-isótropo. Es no-homogéneo porque no todos sus puntos son equivalentes: por ejemplo, su centro geométrico es muy particular en cuanto a su función dinámica. Y es no-isótropo porque no todas las direcciones son equivalentes: por ejemplo, las direcciones radiales son aquéllas en las que se mueven los objetos sublunares en su búsqueda de su lugar natural. El espacio es aquí finito y cerrado, y no es una entidad inerte: el lugar, en tanto punto del espacio y con independencia de los objetos que lo ocupan, ejerce una acción sobre los cuerpos al definir su forma de movimiento. En este universo aristotélico, el ser humano ocupa una posición privilegiada, cerca del centro, que le permite observar y describir con precisión la verdadera dinámica de todo lo real.

*　　*　　*

La idea aristotélica de que la astronomía se explica por la física fue paulatinamente desapareciendo de la filosofía después de Aristóteles. Ya en la Alejandría del siglo II d.C., cuna del posteriormente famoso sistema ptolemaico, física y astronomía eran disciplinas independientes: la astronomía era concebida como una parte de la matemática, dedicada exclusivamente a describir los movimientos de los astros con independencia de sus causas. Y si bien tanto las teorías físicas como las astronómicas fueron variando a través de los siglos, la escisión entre física, dedicada a explicar lo que había sido el mundo sublunar aristotélico, y astronomía, abocada a la descripción de los cielos, perduró incluso hasta los tiempos de Galileo y Kepler: las leyes galileanas se refieren al movimiento de los cuerpos sobre la superficie terrestre, mientras que las leyes de Kepler describen el comportamiento de los planetas en su movimiento, ahora elíptico alrededor del Sol.

Es recién con Newton que se produce la gran unificación entre las leyes de la Tierra y las leyes de los cielos. En sus *Principios Matemáticos de Filosofía Natural* de 1687, Newton demuestra que la caída parabólica de un proyectil sobre la Tierra −descripta por Galileo− no es más que una aproximación de la órbita elíptica −descripta por Kepler− que describiría el proyectil si, al no interponerse la superficie terrestre, pudiera convertirse en un satélite de la Tierra. Por lo tanto, ya no hay dos tipos de leyes, válidos en distintas regiones del universo: en todo el universo, tanto en las cercanías de la Tierra como en los cielos más lejanos, rigen las mismas leyes de movimiento, las leyes de Newton. De este modo se produce un renacimiento de la antigua idea aristotélica de explicar los hechos astronómicos por causas físicas.

Pero esta unificación sólo se obtiene a costa de abandonar por completo la cosmovisión aristotélica. El universo se abre en un espacio infinito e ilimitado, donde no ocupamos el centro por la sencilla razón de que el universo no tiene centro: todos sus puntos y direcciones son equivalentes. Este espacio es además absoluto, pues existe en sí mismo, con independencia de los cuerpos que lo ocupan, y seguiría existiendo aun sin la presencia de cuerpo alguno. El espacio es además una entidad inerte: puede concebirse como un receptáculo para los cuerpos, un escenario infinito donde se desarrollan todos los movimientos. El tiempo, es también absoluto, verdadero y matemático: en sí mismo y por su propia naturaleza fluye uniformemente sin relación con nada externo

En este universo infinito e ilimitado, las leyes de Newton rigen el comportamiento dinámico de los cuerpos en todos sus puntos y en todas sus direcciones. Estas leyes, además, son invariantes en todo sistema de referencia inercial, es decir, en todo sistema de referencia que esté en reposo o en movimiento rectilíneo uniforme respecto del espacio absoluto. Esto significa que no existe ninguna experiencia mecánica que nos permita decidir si nos encontramos en reposo o en movimiento rectilíneo uniforme respecto del espacio absoluto. No obstante, en el universo newtoniano sí podríamos reconocer que nos movemos aceleradamente respecto del espacio absoluto, pues en un sistema de referencia acelerado aparecen fuerzas inerciales que no son producidas por la interacción entre los cuerpos sino que sólo son efecto de la aceleración.

Ya en tiempos de Newton había quienes se oponían al concepto de espacio absoluto que incorporaba su mecánica. El líder de esta oposición fue Leibniz, quien adoptaba una concepción relacional del espacio −y también del tiempo−, según la cual el espacio no existe independientemente de los cuerpos sino que no es más que la red de relaciones entre todos los objetos del universo: no tiene sentido, entonces, pensar en un espacio vacío. La controversia entre el absolutismo de Newton y el relacionalismo de Leibniz se desarrolló a través de una extensa correspondencia entre el propio Leibniz y el teólogo y filósofo Samuel Clarke, vocero de Newton respecto de la cuestión en debate. Tal vez el ejemplo que mejor encarna esta controversia sea el conocido caso del balde en rotación: según Newton, las fuerzas inerciales −centrífugas− que actúan sobre el agua en un balde en rotación manifiestan la aceleración respecto del espacio absoluto; por lo tanto, esta experiencia representa una forma inequívoca de detectar la entidad cuya existencia era puesta en duda por los argumentos de Leibniz.

En definitiva, el espacio newtoniano es homogéneo e isótropo: todos sus puntos y sus direcciones son equivalentes. Hemos dejado de ocupar un lugar central en este universo sin centro: no existe una posición física privilegiada desde donde dar cuenta de lo real. Sin embargo, aún subsiste una perspectiva descriptiva privilegiada desde la cual observar y describir la realidad física, que es la que brinda el espacio absoluto. Es el espacio absoluto el que nos permite definir los sistemas de referencia inerciales donde las leyes de Newton conservan su invariancia y, con ello, nos brinda el punto de vista necesario para describir los verdaderos movimientos de los cuerpos, la verdadera dinámica de la realidad.

<p style="text-align:center">*    *    *</p>

A fines del siglo XIX, el físico y filósofo austríaco Ernst Mach, en su obra *La Ciencia de la Mecánica*, retoma el debate absolutismo versus relacionalismo y, con él, el famoso ejemplo del balde. Adoptando una posición decididamente relacionalista, Mach argumenta que el comportamiento del agua en el balde no prueba la existencia del espacio absoluto, puesto que las mismas fuerzas centrífugas se obtendrían con la rotación del resto de las masas del universo: lo único relevante es el movimiento relativo y, por tanto, rotación del balde y rotación del resto de las masas del universo son sólo dos formas diferentes de describir el mismo fenómeno físico. Si bien en las presentaciones históricas suelen citarse los experimentos de Michelson y Morley como impulsores de la formulación de la Teoría de la Relatividad, Einstein no los consideraba el punto de partida de sus reflexiones; según él mismo reconocería más tarde, fueron las ideas de Mach la mayor influencia para enfrentar la tarea que lo conduciría a la nueva teoría.

Además de la cuestión del espacio absoluto, lo que inicialmente preocupaba a Einstein era la coexistencia de dos teorías supuestamente fundamentales, la Mecánica de Newton y el Electromagnetismo de Maxwell, que parecían difícilmente compatibilizables bajo un mismo marco teórico: la luz −onda electromagnética− respondía a sus propias leyes no-mecánicas, y todos los intentos por explicarla como vibraciones mecánicas de un éter luminífero habían sido infructuosas. Es precisamente este problema el que Einstein encara en su famoso artículo de 1905, "Sobre la Electrodinámica de los Cuerpos en Movimiento", donde se presenta la Teoría Especial de la Relatividad. Allí se formulan, prescindiendo del éter y del espacio absoluto, nuevas leyes de movimiento que rigen el comportamiento no sólo de todo cuerpo con masa, sino también de cualquier onda electromagnética. Al igual que las leyes de Newton, las leyes de la Teoría Especial son invariantes en todo sistema de referencia inercial. Sin embargo, introducen una primera generalización relevante respecto de las leyes newtonianas, puesto que afirman que no existe ninguna experiencia física, no sólo mecánica sino tampoco electromagnética, que nos permita diferenciar el reposo del movimiento rectilíneo uniforme. Pero dado que, además, se ha abandonado el concepto de espacio absoluto, carece de significado hablar de reposo en el mismo sentido en el que hablaba Newton: el concepto de sistema de referencia inercial ya no se define en relación con el espacio absoluto sino que se convierte en un concepto primitivo de la nueva teoría.

Es ya con esta primera teoría de la relatividad que entra en escena una nueva entidad, completamente ajena a cualquier cosmovisión pre-relativista: el espacio-tiempo. El matemático alemán Hermann Minkowski formuló una representación matemática adecuada y fructífera del espacio-tiempo de la Relatividad Especial que, desde entonces, paso a conocerse como 'espacio-tiempo de Minkowski'. Y con el espacio-tiempo, aparecen los resultados más anti-intuitivos de la teoría: la relatividad de los intervalos espaciales y temporales respecto del sistema de referencia considerado. No nos detendremos en estos aspectos de la teoría, pues suelen ser los más abordados en las presentaciones usuales. No obstante, vale la pena recordar aquí que la relatividad en el espacio y en el tiempo no implica en modo alguno que 'todo es relativo': si bien se relativizan ciertas magnitudes consideradas absolutas en teorías anteriores, la Relatividad Especial introduce nuevos absolutos, cuyo ejemplo más inmediato es la velocidad de la luz en el vacío. Pero también aparece una nueva noción de distancia, la distancia espacio-temporal entre dos puntos del espacio-tiempo, que es independiente del sistema de referencia considerado.

Sin embargo, a pesar de las novedades conceptuales introducidas por la teoría, el problema de Mach seguía sin haber sido completamente resuelto. En efecto, las leyes de la Relatividad Especial aún distinguen entre sistemas de referencia inerciales y no-inerciales; en consecuencia, la aceleración parece conservar un cierto carácter absoluto incompatible con el relacionalismo de Mach. Con el objetivo de satisfacer las exigencias machianas, Einstein se propone generalizar su Relatividad Especial; el resultado de esta tarea es la famosa Teoría General de la Relatividad, cuya versión definitiva aparece en 1916.

En la Relatividad General ya no hay sistemas de referencia privilegiados: las leyes son invariantes en todo sistema de referencia. Pero esto se logra a costa de dotar al espacio-tiempo de una propiedad muy peculiar, como la de curvarse ante la presencia de cuerpos con masa. En este nuevo marco conceptual, la fuerza gravitatoria desaparece de la escena: los cuerpos ya no se mueven como consecuencia de su interacción gravitatoria con otros cuerpos, sino que lo hacen siguiendo el 'camino' más corto –la geodésica– sobre un espacio-tiempo curvo.

Es precisamente la naturaleza del espacio-tiempo el aspecto que más ha preocupado a los filósofos de la física desde la formulación de la Teoría General de la Relatividad. ¿Qué tipo de entidad es este espacio-tiempo que puede deformarse debido a los cuerpos con masa que lo ocupan?, ¿es el espacio-tiempo una sustancia? En algún sentido parece serlo, pues ya no se asimila a un receptáculo inerte para los cuerpos: al igual que las sustancias, posee propiedades definidas en cada uno de sus puntos –su curvatura–. No obstante, aun si pudiera pensarse como una sustancia, el espacio-tiempo sería una sustancia completamente excepcional, pues no establece sus relaciones con las demás sustancias a través de una interacción análoga a las interacciones físicas conocidas.

En cierto sentido, el espacio-tiempo de Einstein tiene más puntos de contacto con el espacio aristotélico que con el espacio absoluto newtoniano. En general, el espacio-tiempo de la relatividad general, al igual que el espacio de Aristóteles, no es homogéneo ni isótropo puesto que su curvatura varía en cada uno de sus puntos. Por otra parte, como en el caso del espacio aristotélico, no es una entidad inerte, totalmente pasiva: el espacio-tiempo ejerce una muy particular acción sobre los cuerpos en la medida en que éstos modifican su estado de movimiento frente a la curvatura del espacio-tiempo. Además, las ecuaciones de campo de Einstein admiten como soluciones modelos de universos cerrados. En estos casos puede repetirse, en una versión espacio-temporal, lo dicho en la descripción del universo de Aristóteles: el universo de la Relatividad General es todo aquello que existe en el espacio-tiempo; 'fuera' del espacio-tiempo no hay nada, ni materia, ni espacio, ni tiempo. Poco queda entonces de la idea newtoniana de un receptáculo infinito e ilimitado, homogéneo, isótropo e inerte, que juega el papel de escenario pasivo donde los cuerpos se mueven e interactúan.

Estas similitudes entre el universo de Aristóteles y el universo de Einstein no debe, sin embargo, ocultar una diferencia central: mientras que el universo aristotélico se piensa en el espacio y en el tiempo, la Relatividad General incorpora el tiempo a una estructura de cuatro dimensiones donde las tres dimensiones espaciales y la dimensión temporal, si bien distinguibles entre sí, se encuentran estrechamente 'entrelazadas'. Por ello, tal vez sea el concepto de tiempo el que sufre una mutación más drástica respecto de la cosmovisión newtoniana. El tiempo ya no es aquello que fluye uniformemente, sin relación con nada externo, de lo cual nos hablaba Newton. Como muchos autores señalan, la profunda diferencia entre los conceptos de tiempo en Relatividad General y en las teorías físicas pre-relativistas es uno de los principales obstáculos para la unificación entre relatividad y cuántica. En la Teoría Cuántica, el tiempo es el parámetro de evolución de un sistema y, por tanto, 'externo' al sistema mismo, lineal y absoluto. En Relatividad General, por el contrario, el tiempo se convierte en una dimensión de una variedad cuatridimensional que se curva a gran escala frente a la presencia de masas. Y aún no se ha logrado subsumir ambas teorías, con sus respectivos conceptos de tiempo, bajo un único marco teórico más general.

Pero volvamos ahora al comienzo de nuestra historia acerca de la Teoría de la Relatividad: ¿logra finalmente Einstein, con su Relatividad General, satisfacer el relacionalismo de Mach? Si bien en la actualidad algunos autores todavía sueñan con una reconstrucción totalmente relacional de la teoría, en general los filósofos de la física concuerdan en admitir que la Relatividad General no satisface por completo las exigencias machianas. El principal obstáculo para concebir la Relatividad General como una teoría relacionalista respecto del espacio-tiempo reside en el hecho de que las ecuaciones de campo tienen solución para el caso de un espacio-tiempo vacío: mientras que para una concepción relacionalista carece de sentido hablar de un espacio-tiempo vacío, en el marco teórico de la Relatividad General el espacio-tiempo vacío existe, y es precisamente el espacio-tiempo plano de Minkowski.

<div align="center">*     *     *</div>

Este recorrido a través del devenir del pensamiento filosófico nos permite comprender cómo ha ido modificándose, a través de las tres grandes cosmovisiones de la historia, nuestra concepción del universo y de la ubicación física y gnoseológica que ocupamos en él.  En particular, podemos comprobar el progresivo descentramiento que ha sufrido el sujeto en cuanto a su posición en el cosmos y a su conocimiento de lo real.  En el universo aristotélico, esférico y cerrado, ocupamos una posición central y única que nos permite describir la verdadera dinámica de la realidad; realidad que podemos conocer en sí misma por ser isomórfica con nuestro lenguaje y nuestro pensamiento. En el universo newtoniano, eterno y sin centro, perdemos definitivamente nuestra posición física privilegiada.   No obstante, conservamos aún la capacidad de conocer la realidad desde una perspectiva gnoseológica privilegiada, que es la que nos brinda el espacio absoluto.  Finalmente, en la cosmovisión einsteniana el descentramiento del sujeto se profundiza al extenderse también al plano gnoseológico.   En efecto, no sólo hemos perdido la centralidad que nos adjudicaban las cosmologías geocéntricas, sino que ya no existe un punto de vista privilegiado desde donde dar cuenta de lo real.   En tanto seres no-espacio-temporales sino espaciales y temporales, nuestra perspectiva descriptiva está siempre ligada a un sistema de referencia.   Pero dado que no existe sistema de referencia privilegiado, todas las perspectivas son igualmente legítimas: nuestro acceso empírico a la realidad es siempre y necesariamente perspectival.  Ya nada queda, entonces, de aquel sujeto clásico que, desde su posición privilegiada, podía acceder a lo real y brindar su única descripción verdadera.

<div align="center">*     *     *</div>

Hasta aquí hemos adoptado el punto de vista de las teorías físicas para reflexionar acerca de los universos que ellas nos describen.  Pero, ¿qué nos ha enseñado la Teoría de la Relatividad en un plano epistemológico metateórico?  Como es bien sabido, después de más de dos siglos de reinado absoluto de la Mecánica Newtoniana, cuando muchos físicos de fines del siglo XIX creían seriamente que la física había alcanzado sus confines teóricos, en las primeras décadas del siglo XX el ámbito científico se vio sacudido por dos teorías, la Mecánica Cuántica y la Teoría de la Relatividad, que modificaron por completo el conocimiento físico.

Esta revolución en la física tuvo una enorme influencia sobre el pensamiento epistemológico de la época, poniendo en crisis un empirismo-inductivismo decimonónico según el cual la ciencia avanza de un modo acumulativo, agregando conocimiento a los conocimientos definitivos previamente alcanzados. Frente a los nuevos acontecimientos en física, los epistemólogos rápidamente advirtieron que la ciencia no progresa por mera acumulación, sino que los verdaderos hitos de su desarrollo vienen dados por el reemplazo de teorías. Pero, además, fue reconociéndose que, más allá de sus resultados específicos, cada teoría conlleva su propia visión de la realidad, su propia ontología: el cambio de teoría no conduce a una mera redescripción de las mismas entidades que poblaban el mundo descripto por la teoría previa; la nueva teoría nos coloca en un nuevo mundo, en una nueva realidad con sus propias entidades, propiedades y relaciones. La ontología de la Relatividad General es completamente diferente de la ontología newtoniana: ambos mundos son incluso inconmensurables.

Si el devenir de la ciencia nos enfrenta a ontologías permanentemente cambiantes en cuanto a sus estructuras más básicas, ya no hay razón para pensar que somos capaces de conocer la realidad tal como es en sí misma, con independencia de los esquemas conceptuales que utilizamos para describirla. Aun admitiendo que el conocimiento científico se desarrolla hacia la formulación de teorías más adecuadas y fructíferas, ya son pocos los filósofos que continúan creyendo que la ciencia avanza hacia la teoría definitiva y verdadera acerca de lo real en sí. Una vez que hemos renunciado a describir la realidad desde el punto de vista del 'ojo de Dios', nuestro conocimiento se convierte en un saber acotado y profundamente humano. Y si bien para algunos esta posición puede conducir al desengaño, el abandono de las verdades absolutas y definitivas y la idea de un conocimiento provisorio y revisable son excelentes antídotos contra los dogmatismos que suelen poner de manifiesto los aspectos más oscuros del ser humano. También en el sentido de promover un conocimiento humanizado y abierto al disenso, la Teoría de la Relatividad ha contribuido de un modo fundamental.